\documentclass[aps,prb,twocolumn, showpacs]{revtex4}

\usepackage{epsfig}
\usepackage{graphicx}

\def\be{\begin{equation}}
\def\ee{\end{equation}}
\def\bea{\begin{eqnarray}}
\def\eea{\end{eqnarray}}

\begin{document}
\draft

\title{Sum rules for spin-Hall conductivity cancelation}
\author{Emmanuel I. Rashba\cite{Rashba*}}
\affiliation{Department of Physics, MIT, Cambridge, Massachusetts 02139, USA}
\date{September 14, 2004}

\begin{abstract}
It has been shown recently that the universal dc spin conductivity of two-dimensional electrons with a Rashba spin-orbit interaction is canceled by vertex corrections in a weak scattering regime. We prove that the zero bulk spin conductivity is an intrinsic property of the free-electron Hamiltonian and scattering is merely a tool to reveal this property in terms of the diagrammatic technique. When Zeeman energy is neglected, the zero dc conductivity persists in a magnetic field. Spin conductivity increases resonantly at the cyclotron frequency and then decays towards the universal value.
\end{abstract}

\maketitle

\narrowtext
In the framework of the burgeoning field of semiconductor spintronics,\cite{Wolf,ALS02,ZFDS04} an active interest currently exists in the mechanism of generating spin fluxes by electric fields. Pure spin fluxes that are not accompanied by electric currents are expected to reduce dissipative losses related to spin injection and spin transport. Therefore, the proposals by Murakami {\it et al.}\cite{Mur} and Sinova {\it et al.}\cite{Sino} of generating dissipationless tranverse spin currents (a spin-Hall effect) by a driving dc electric field acquired a lot of attention. For two-dimensional (2D) systems with a spin-split (Rashba) energy spectrum the universal spin conductivity of Ref.~\onlinecite{Sino} has been put under thorough scrutiny. It turned out that the dissipationless component of the bulk spin current, which originates from virtual interbranch transitions, is canceled by ladder diagrams when a weak impurity scattering is taken into account. This cancelation was originally discovered by Schwab and Raimondi\cite{SR02} in a somewhat different context, rediscovered by Inoue {\it et al.},\cite{IBM04} and recently confirmed in a number of papers.\cite{MSH04,Kh04,RS04,Dim} The cancelation is rather puzzling because it does not follow from any known symmetry arguments. Therefore, it is important to uncover its origin.

The Hamiltonian in question is
\be
{\hat H}=\hbar^2{\hat{\mbox{\boldmath$k$}}}^2/2m+\alpha(\mbox{\boldmath$\sigma$}\times\hat{\mbox{\boldmath$k$}})\cdot\hat{\bf z},
\label{eq1}
\ee
and in the absence of magnetic field, $\mbox{\boldmath$B$}=0$, spin currents are defined as
\be
{\cal J}_{zj}={1\over 2}\sum_\lambda\int{{d^2k}\over(2\pi)^2}\langle\lambda\vert\sigma_z{\hat v}_j({\mbox{\boldmath$k$}})+{\hat v}_j({\mbox{\boldmath$k$}})\sigma_z\vert\lambda\rangle
\label{eq2}
\ee
with a proper choice of the integration area in the $\mbox{\boldmath$k$}$-space. Here $\hat{\mbox{\boldmath$v$}}=\hbar^{-1}\partial{\hat H}/\partial{\mbox{\boldmath$k$}}$ is the velocity operator, $j=x,y$, and $\hat{\bf z}$ is a unit vector perpendicular to the confinement plane. Spin currents ${\cal J}_{zx}$ and ${\cal J}_{zy}$ are driven by electric fields $E_y$ and $E_x$, respectively. The corresponding spin conductivities are $\Sigma_{zxy}={\cal J}_{zx}/E_y$ and $\Sigma_{zyx}={\cal J}_{zy}/E_x$. Because the Hamiltonian $\hat H$ possesses a $\mbox{\boldmath$C$}_{\infty v}$ symmetry, applying $\sigma_v$ reflections results in $\Sigma_{zxy}=-\Sigma_{zyx}.$ Hence, the part of the tensor $\Sigma_{zij}$ related to the two last indices is equivalent to $M_z$, a $z$ component of a pseudovector $\mbox{\boldmath$M$}$ orthogonal to the $xy$ plane, and a product $\sigma_zM_z$ is an invariant of the group $\mbox{\boldmath$C$}_{\infty v}$. These arguments prove that $\Sigma_{zij}$ is an invariant tensor of the spatial group. It also invariant with respect to the $t$-inversion. Under these conditions, one can expect that vanishing of $\Sigma_{zij}$ is a result of some sum rules, and we present these sum rules in what follows.

The oscillator strength of a free electron state related to the ``transition to itself" manifests itself in the absorption as a cyclotron or Drude spectrum, i.e., only in the presence of a perturbation like a magnetic field or some electron scattering mechanism. The same problem can be relevant for spin currents, hence, we impose a magnetic field ${\mbox{\boldmath$B$}}\parallel{\hat{\bf z}}$ by considering $\hat{\mbox{\boldmath$k$}}$ in Eq.~(\ref{eq1}) as a kinetic momentum $\hat{\mbox{\boldmath$k$}}=-i\mbox{\boldmath$\nabla$}+e\mbox{\boldmath$A$}/\hbar c$, but do not include a Zeeman term into the model Hamiltonian $\hat H$. The exact solution of this problem is\cite{R60}
\be
\Psi_{\lambda n}=\left(
\begin{array}{c}
i\lambda b_n^{\bar\lambda}\psi_{n-1}\\b_n^\lambda\psi_n
\end{array}\right),~~E_{\lambda n}=\hbar\omega_c(n+\lambda c_n)
\label{eq3}
\ee
where
\be
b_n^\lambda={1\over{\sqrt2}}\sqrt{1+{\lambda/{2c_n}}},
c_n=\sqrt{\gamma^2n+{1/4}},{\bar\lambda}=-\lambda.
\label{eq4}
\ee
Here $\gamma=[{2(m\alpha^2/\hbar^2)}/{\hbar\omega_c}]^{1/2}$ is a dimensionless spin-orbit coupling constant, $\hbar\omega_c=eB/\hbar c$ is the cyclotron frequency, $\lambda=\pm1$ designates the spectrum branches, $n\geq1$, and $\psi_n$ are oscillator wave functions. The eigenstate $n=0$ exists only for $\lambda=1$, with $b_0^1=1$ and $b_0^{\bar 1}=0$.

The standard Kubo-Greenwood formula results in the following expression for the frequency dependence of the real part of spin conductivity 
\be
\Sigma_{zxy}(\omega)=-{{ie}\over{\pi\hbar{\it l}^2}}\sum_{\lambda n\lambda'n'}^{*}
{{\langle\lambda n\vert{\hat v}_y\vert\lambda'n'\rangle\langle\lambda'n'\vert{\hat{\cal J}}_{zx}\vert\lambda n\rangle}\over{(\omega_{\lambda' n'}-\omega_{\lambda n})^2-\omega^2}}.
\label{eq5}
\ee
The asterisk over the summation sign indicates that summation should be performed only over the states with $\omega_{\lambda n}\equiv E_{\lambda n}/\hbar>\mu/\hbar$, $\omega_{\lambda'n'}<\mu/\hbar$, $\mu$ being the Fermi energy; below, we accept $\mu>0$. Here ${\it l}=(c\hbar/eB)^{1/2}$ is the magnetic length. It appeared through the factor $1/2\pi {\it l}^2$ accounting for the Landau level degeneracy. When deriving Eq.~(\ref{eq5}) we took into account that the matrices $\langle\lambda n\vert{\hat v}_y\vert\lambda'n'\rangle$ and $\langle\lambda n\vert{\hat{\cal J}}_{zx}\vert\lambda' n'\rangle$ are antisymmetric and symmetric, respectively. For the same reason, the sum includes only nondiagonal matrix elements.

It is seen from Fig.~1 that for $\lambda\neq\lambda'$ the summation over $n$ should be performed over a vast area, while for $\lambda=\lambda'$ the sum includes, for each branch, only a single term coming from the two states adjacent to the Fermi level. It will be shown in what follows that {\it these two terms are anomalously large and in the dc regime cancel the sum}.

To simplify Eq.~(\ref{eq5}), one can take advantage of the fact that both operators in the numerator can be expressed through commutators including the Hamiltonian $\hat H$. First, ${\hat v}_y=(i/\hbar)[{\hat H},{\hat y}]_-$, where the operator of the coordinate ${\hat y}={\hat y}_0-{\it l}^2{\hat k}_x$, and ${\hat y}_0$ is the center-of-orbit operator.\cite{JL49} The operators ${\hat k}_x$ and ${\hat k}_y$ are related, in the standard way, to the creation and annihilation operators, ${\hat k}_x=(a^++a)/{\it l}\sqrt{2}$ and ${\hat k}_y=(a^+-a)/i{\it l}\sqrt{2}$. Because ${\hat y}_0$ commutes with $a$ and $a^+$, it can be disregarded in nondiagonal matrix elements of Eq.~(\ref{eq5}), hence,
\be
\langle\lambda n\vert{\hat v}_y\vert\lambda'n'\rangle=-i{\it l}^2(\omega_{\lambda n}-\omega_{\lambda'n'})
\langle\lambda n\vert{\hat k}_x\vert\lambda'n'\rangle.
\label{eq6}
\ee
Second, the spin-current operator ${\cal J}_{zx}$ equals ${\hat{\cal J}}_{zx}=-(i\hbar/2m\alpha)[{\hat H},\sigma_x]_-$ as has been noticed by Dimitrova.\cite{Dim} Therefore
\be
\langle\lambda' n'\vert{\hat{\cal J}}_{zx}\vert\lambda n\rangle=-{{i\hbar^2}\over{2m\alpha}}(\omega_{\lambda' n'}-\omega_{\lambda n})
\langle\lambda' n'\vert\sigma_x\vert\lambda n\rangle.
\label{eq7}
\ee
After substituting Eqs.~(\ref{eq6}) and (\ref{eq7}) into Eq.~(\ref{eq5}), we arrive at a simple equation for $\Sigma_{zxy}\equiv\Sigma_{zxy}(\omega=0)$
\be
\Sigma_{zxy}
=-{{ie\hbar}\over{2\pi m\alpha}}\sum_{\lambda n\lambda'n'}^*
\langle\lambda n\vert{\hat k}_x\vert\lambda'n'\rangle
\langle\lambda'n'\vert\sigma_x\vert\lambda n\rangle.
\label{eq8}
\ee

It is seen from Fig.~1 that there are two types of transitions that contribute to the sum of Eq.~(\ref{eq8}). Interbranch transitions from the upper ($\lambda=1$) branch to the lower ($\lambda=-1$) branch are possible everywhere inside the shadowed strips. They obey the oscillator selection rules $(1,n)\rightarrow({\bar 1},n\pm1$). Intrabranch transitions are possible only in the immediate vicinity of the Fermi level, from the lower empty orbit to the upper filled orbit.

The explicit expression for the contribution of interbranch transitions to $\Sigma_{zxy}$ is
\be
\Sigma_{zxy}^{1{\bar 1}}={{e\hbar}\over{8\pi\sqrt{2}m\alpha{\it l}}}
\sum_{n\nu}^*\gamma{{(c_n+\nu/2)+\nu n(c_n-c_{n+\nu})}\over{c_nc_{n+\nu}}}
\label{eq10}
\ee
where the terms with $\nu=\pm1$ describe the contributions from $(1,n)\rightarrow({\bar 1},n\pm1)$ transitions, respectively. For weak magnetic fields, when $\gamma,n\gg1$, the separate terms of the sum take asymptotic values $1/2\sqrt{n}$, and summation can be replaced by integration over $n$. Phase volume arguments suggest the relation $n_\pm={\it l}^2k_\pm^2/2$ between the integration boundaries $n_\pm$ and the Fermi momenta $k_\pm$ of the spectrum branches. Finally
\be
\Sigma_{zxy}^{1{\bar 1}}={{e\hbar}\over{8\pi\sqrt{2}m\alpha{\it l}}}
\int_{n_+}^{n_-}{{dn}\over{\sqrt{n}}}={{e\hbar(k_--k_+)}\over{8\pi m\alpha}}
={e\over{4\pi\hbar}}
\label{eq11}
\ee
because $k_--k_+=2m\alpha/\hbar^2$. This result coincides with the universal conductivity by Sinova {\it et al.}\cite{Sino} (with the accuracy to the factor $\hbar/2$ due to the difference in the definition of ${\hat{\cal J}}_{zx}$).

Equation (\ref{eq11}) indicates that the interband contribution to $\Sigma_{zxy}$, that includes a sum over the whole ``bulk" of the Fermi sea, in the $B\rightarrow0$ limit is expressed in terms of the Fermi surface parameters, the momenta $k_-$ and $k_+$. This result resembles the recent inference by Haldane\cite{H04} that the anomalous Hall effect is a Fermi surface property.

Two intrabranch contributions to $\Sigma_{zxy}$ come from the diagonal in $\lambda$ terms in Eq.~(\ref{eq8})
\bea
\Sigma^{\lambda\lambda}_{zxy}&=&-{{ie\hbar}\over{2\pi m\alpha}}
\langle\lambda n_\lambda\vert{\hat k}_x\vert\lambda n_\lambda-1\rangle
\langle\lambda n_\lambda-1\vert\sigma_x\vert\lambda n_\lambda\rangle\nonumber\\
&&\nonumber\\
&=&{{\lambda e\hbar\gamma}\over{8\pi\sqrt{2}m\alpha{\it l}}}
{{(n_\lambda-1)c_{n_\lambda}+n_\lambda c_{n_\lambda-1}+\lambda/2}\over{c_{n_\lambda}c_{n_\lambda-1}}}.
\label{eq12}
\eea
Here $n_\lambda$ is the lower empty orbit at the $\lambda$ branch, $\lambda=\pm1$, hence, these contributions are Fermi surface parameters. For a weak magnetic field, Eq.~(\ref{eq12}) reduces to
\be
\Sigma^{\lambda\lambda}_{zxy}=\lambda{{e\hbar k_\lambda}\over{8\pi m\alpha}}.
\label{eq13}
\ee  
Therefore, in the $B\rightarrow0$ limit,
\be
\Sigma_{zxy}=\Sigma^{1{\bar 1}}_{zxy}+\Sigma^{11}_{zxy}+\Sigma^{{\bar 1}{\bar 1}}_{zxy}=0,
\label{eq14}
\ee
in agreement with the conclusions of Refs.~\onlinecite{SR02,IBM04,MSH04,Kh04,RS04}. We note that two intraband terms $\Sigma^{\lambda\lambda}_{zxy}$ could cancel the sum of Eq.~(\ref{eq10}) because they are large as $n^{1/2}$ while the separate terms of the sum are small as $n^{-1/2}$.

The above derivation of the zero spin-conductivity, $\Sigma_{zxy}=0$, is valid in the $\omega/\omega_c\rightarrow0, \omega_c\rightarrow0$ limit and does not depend on any assumptions related to the scattering mechanism (provided the scattering is weak) and is applicable for an arbitrary ratio of the spin-orbit coupling energy $m\alpha^2/\hbar^2$ to $\mu$. This derivation of the cancelation theorem provides a direct connection to the interbranch spin conductivity of Eq.~(\ref{eq11}) that has been found and generalized in a number of papers.\cite{Sino,SL04,Ch04,Sin04,SMXZ,R04,ESL04,D04} In what follows we provide a formal proof that shows that this theorem is valid at arbitrary $B$ and follows from two sum rules.

The first sum rule follows from the fact that the matrices $\langle\lambda n\vert{\hat k}_x\vert\lambda'n'\rangle$ and $\langle\lambda n\vert\sigma_x\vert\lambda' n'\rangle$ are symmetric and antisymmetric, respectively. Therefore, the numerator of Eq.~(\ref{eq8}) can be rewritten through a commutator of two operators, ${\hat k}_x$ and $\sigma_x$, that commute. For a sum over $\lambda'n'$ (restricted only by the oscillator selection rule $n'=n\pm1$) we arrive at the identity
\be
\sum_{\lambda'n'}
\langle\lambda n\vert{\hat k}_x\vert\lambda'n'\rangle
\langle\lambda'n'\vert\sigma_x\vert\lambda n\rangle={1\over2}
\langle\lambda n\vert[{\hat k}_x,\sigma_x]_-\vert\lambda n\rangle=0
\label{eq14a}
\ee
that is valid for arbitrary $(\lambda, n)$. The second identity reads
\be
\sum_{\lambda'}\langle\lambda n\vert {\hat k}_x\vert\lambda'n+\lambda\rangle\langle\lambda'n+\lambda\vert\sigma_x\vert\lambda n\rangle=-2n/c_n.
\label{eq14b}
\ee
It is important that this sum does not depend on $\lambda$. The latter identity can be checked by inspection using the eigenfunctions of Eq.~(\ref{eq3}). 

To calculate $\Sigma_{zxy}$ of Eq.~(\ref{eq8}), we introduce an auxiliary sum
\be
\Sigma_{zxy}^0\equiv
-{{ie\hbar}\over{2\pi m\alpha}}\sum_{n_+\leq n\leq n_-}\sum_{\lambda'n'}
\langle 1 n\vert{\hat k}_x\vert\lambda'n'\rangle
\langle\lambda'n'\vert\sigma_x\vert 1 n\rangle
\label{eq15}
\ee
with $n$ running across the shadowed strips of Fig.~1 and arbitrary $\lambda'$ and $n'$. The interbranch part of the sum, $\lambda'=-1$, differs from $\Sigma_{zxy}^{1{\bar 1}}$ only by two extra terms with $n'$ near $n_-$. The intrabranch part of the sum, $\lambda'=1$, highly simplifies because the summand in Eq.~(\ref{eq15}) is antisymmetric in the quantum numbers, $1n$ and $1n'$. As a result, all terms but two, the last on the left and on the right, cancel. Finally, the equation relating $\Sigma_{zxy}$  and $\Sigma_{zxy}^0$ reads
\widetext
\bea
\Sigma_{zxy}=\Sigma_{zxy}^0
+{{ie\hbar}\over{2\pi m\alpha}}\biggl( \langle 1 n_--1\vert{\hat k}_x\vert{\bar 1}n_-\rangle\langle{\bar 1}n_-\vert\sigma_x\vert 1 n_--1\rangle
&+&\langle 1 n_-\vert{\hat k}_x\vert {\bar1}n_-+1\rangle\langle {\bar1}n_-+1\vert\sigma_x\vert 1 n_-\rangle\nonumber\\
+\langle 1 n_-\vert{\hat k}_x\vert1n_-+1\rangle\langle1n_-+1\vert\sigma_x\vert 1 n_-\rangle
&-&\langle {\bar1} n_-\vert{\hat k}_x\vert{\bar1}n_--1\rangle\langle{\bar1}n_--1\vert\sigma_x\vert {\bar1} n_-\rangle\biggr).
\label{eq15a}
\eea
\endwidetext 
\noindent
In the parentheses, the two first terms are interbranch contributions and the two last terms are intrabranch contributions. 

Now we are in a position to apply the sum rules of Eqs.~(\ref{eq14a}) and (\ref{eq14b}) to Eq.~(\ref{eq15a}). $\Sigma_{zxy}^0$ vanishes due to Eq.~(\ref{eq14a}). The second and third terms in the parentheses can be summed by using Eq.~(\ref{eq14b}). After transforming the first term by using the symmetry properties of the matrices ${\hat k}_x$ and $\sigma_x$, the first and the fourth terms in the parentheses can be also summed, and the two sums cancel. Finally, we arrive at the cancelation theorem, $\Sigma_{zxy}=0$, valid for arbitrary $B$.

We conclude that {\it the cancelation theorem of Eq.~(\ref{eq14}) is a corollary of the sum rules of Eq.~(\ref{eq14a}) and (\ref{eq14b}).} It is an exact property of the free-electron Hamiltonian $\hat H$, and spin conductivity should increase gradually with increasing electron scattering as has been found numerically in Ref.~\onlinecite{RS04}. Therefore, dc spin conductivity of the systems described by the Hamiltonian of Eq.~(\ref{eq1}) is dissipative in agreement with the conclusions of Refs.~\onlinecite{IBM04,MSH04,Kh04,RS04}. The differences in the results of independent numerical efforts\cite{RS04,XX04,Nomura,SST04} are not properly understood yet and can originate from the differences in boundary conditions.

It is instructive to follow up how the cancelation rule of Eq.~(\ref{eq14}) breaks down in ac regime. When $\omega\neq0$, each term in Eqs.~(\ref{eq10}) and (\ref{eq12}) should be multiplied by a factor
\be
f_{\lambda n\lambda'n'}(\omega)={{(\omega_{\lambda n}-\omega_{\lambda'n'})^2}\over{(\omega_{\lambda n}-\omega_{\lambda'n'})^2-\omega^2}}.
\label{eq16}
\ee
When $\omega\ll2\alpha k_\pm/\hbar$, these factors are close to unity for interbranch transitions, hence, the finite-frequency correction to $\Sigma^{1{\bar 1}}_{zxy}$ can be disregarded. On the contrary, intrabranch terms $\Sigma^{\lambda\lambda}_{zxy}(\omega)$ change strongly at the scale of $\omega_c$. When $\omega\ll\omega_c$, the relative corrections to the sum $\Sigma^{11}_{zxy}+\Sigma^{{\bar 1}{\bar 1}}_{zxy}$ are about $3\omega^2/\omega_c^2$. The sum $\Sigma^{11}_{zxy}(\omega)+\Sigma^{{\bar 1}{\bar 1}}_{zxy}(\omega)$ diverges at $\omega\approx\omega_c$, changes sign, and decreases as $(\omega_c/\omega)^2$ for $\omega\gg\omega_c$. In the frequency region $\omega_c\ll\omega\ll2\alpha k_\pm/\hbar$, spin conductivity
$\Sigma_{zxy}(\omega)\approx\Sigma^{1{\bar 1}}_{zxy}$, hence, ac spin conductivity approaches the universal value.\cite{Sino} This conclusion correlates with the results by Inoue {\it et al.}\cite{IBM04} and Mishchenko {\it et al.}\cite{MSH04} who have found that in clean samples spin conductivity increases at $\omega\sim\hbar/\tau_p$, $\tau_p$ being the momentum relaxation time. All these results emphasize the importance of transients for generating electrically driven spin currents.\cite{R04,R04SB}

The above results have been derived for a homogeneous system in a homogeneous electric field. Inhomogeneities facilitate spin currents (and spin injection) both in the diffusive\cite{MSH04,BNMD03} and ballistic regimes.\cite{R04,R04SB} Mishchenko {\it et al.}\cite{MSH04} have predicted developing transport spin currents near the electrodes, within a spin diffusion length from them, and spin accumulation near the corners. On the contrary, equilibrium edge spin currents studied by Reynoso {\it et al.}\cite{RUSB04} are background currents\cite{R03} and do not produce spin accumulation. Spin currents near the free edges parallel to the field $\mbox{\boldmath$E$}\parallel{\hat{\bf y}}$ have not been investigated yet. In general, establishing a connection between spin currents and spin accumulation is a demanding and important problem. Indeed, despite the fact that transport spin currents can, in principle, be measured (because any deviation of an electronic system from equilibrium produces electro-motive forces\cite{LL36}), and some procedures have been already proposed,\cite{Malsh} it is the spin magnetization that is the principal measurable quantity.

The cancelation theorem of Eq.~(\ref{eq14}) is valid also for a Dresselhaus Hamiltonian\cite{D55,PP95} because it differs from the Rashba Hamiltonian only by an unitary transformation $\sigma_x\rightarrow\sigma_y,\sigma_y\rightarrow\sigma_x,\sigma_z\rightarrow-\sigma_z$ that does not change Eq.~(\ref{eq5}).

For comparison, the Drude formula for Hall conductivity $\sigma_{xy}=(e^2n\omega_c\tau_p^2/m)/(1+\omega_c^2\tau_p^2)$ has different $\omega_c\rightarrow0$ and $\tau_p^{-1}\rightarrow0$ limits, $\sigma_{xy}=0$ and $\sigma_{xy}=e^2n/m\omega_c$, respectively. On the contrary, the diagrammatic ($\omega_c=0$) and our ($\tau_p^{-1}=0$) results for dc spin-Hall conductivity match smoothly, what indicates a remarkable stability of its vanishing.

In conclusion, we have shown that the cancelation of the inter- and intrabranch contributions to the dc bulk spin conductivity ${\cal J}_{zxy}$ is an intrinsic property of the free electron Hamiltonian and originates from two sum rules inherent in it. Weak electron scattering is only needed to reveal this cancelation in terms of the diagrammatic technique. The cancelation persists in a magnetic field if the Zeeman interaction is disregarded, and spin conductivity increases resonantly at the cyclotron frequency.
 
I am grateful to O. V. Dimitrova and M. V. Feigel'man for stimulating correspondence. Funding of this research granted through a DARPA contract is gratefully acknowledged.

%\noindent

{\bf Figure Captions}

FIG.~1. Schematic of the energy spectrum and electronic transitions in a magnetic field. Here $\lambda=\pm1$ numerates spectrum branches, $k_+$ and $k_-$ are Fermi momenta for the upper and lower branches, respectively, $n$ are Landau quantum numbers for these branches, $n_\pm$ are the lowest unoccupied states at both branches, and $\mu$ is the Fermi level. Shadowed strips indicate regions of interbranch transitions between the occupied $(-)$-states and unoccupied $(+)$-states. The arrows $n\rightarrow n\pm1$ show typical interbranch transitions in the ``bulk" of the Fermi sea, while the arrows $n_+\rightarrow n_+-1$ and $n_-\rightarrow n_--1$ indicate two intrabranch transitions at the Fermi edge. Cancelation of the inter- and intrabranch contributions results in the zero bulk spin conductivity ${\cal J}_{zxy}$.


\begin{thebibliography}{99}
\bibitem{gfactor} This conclusion has been derived for the Hamiltonian $\hat H$ with a zero $g$-factor. It is exact in the $B=0$ limit, and at $B\neq0$ is applicable to the AlGaAs alloys with $g\approx0$.\cite{Kato03} The case of large $g$-factors needs a special consideration.
\end{thebibliography}
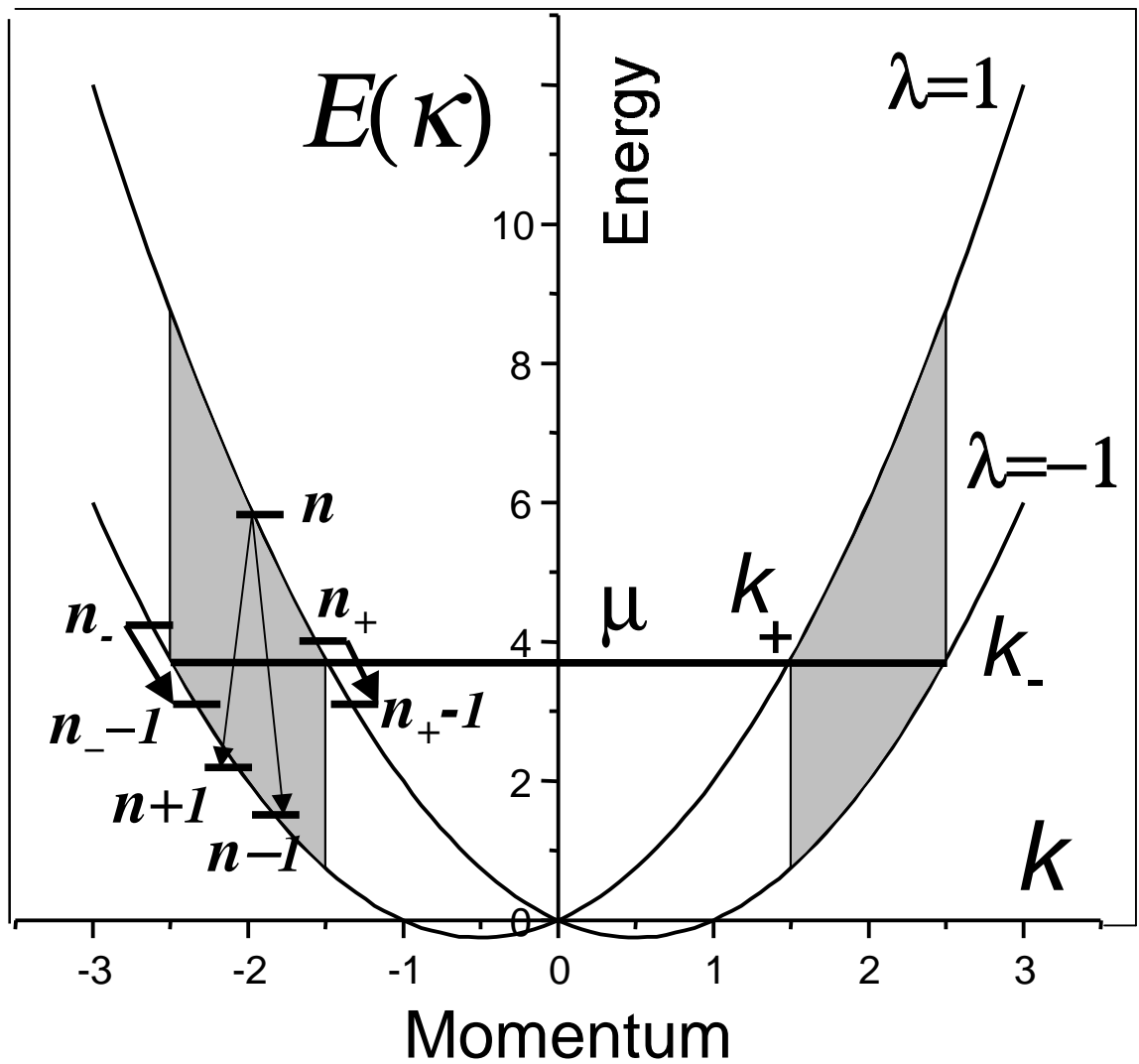
\end{document}